\documentclass[aps,prc,preprint,groupedaddress,nofootinbib]{revtex4-1}

\usepackage{graphicx}
\usepackage{units}
\usepackage{amssymb}
\usepackage{amsmath}
\usepackage{hyperref}

 \newcommand\la{\langle}
 \newcommand\ra{\rangle}
 \newcommand\beq{\begin{equation}}
 
 \newcommand\eeq{\end{equation}}
 \newcommand\beqn{\begin{eqnarray}}
 \newcommand\eeqn{\end{eqnarray}}

\begin{document}

\title{Rapidity Distributions of Hadrons in Proton-Nucleus Collisions}

\author{H. J. Pirner$^{1}$}
\author{B. Z. Kopeliovich$^2$}

\affiliation{\centerline
{$^{1}$Institute for Theoretical Physics, University of Heidelberg, Germany}\\
{$^2$Departamento de F\'{\i}sica,
Universidad T\'ecnica Federico Santa Mar\'{\i}a; and}
Centro Cient\'ifico-Tecnol\'ogico de Valpara\'iso;
Casilla 110-V, Valpara\'iso, Chile}

\begin{abstract}
We study proton-lead collisions with a new model for the Fock states of the incoming proton. The number of collisions, which the proton experiences, selects the appropriate Fock state of the proton which generates a multiple of pp-like rapidity distributions. We take as input the pp maximum entropy distributions shifting the respective cm-rapidities and reducing the available energies. A comparison with existing data at 5 TeV is made, and results for 8 TeV are presented. We also explore the high multiplicity data in this model.
\end{abstract}



\maketitle



\section{Introduction}
\label{sec:intro}

The main ideas underlying high energy proton-nucleus collisions are well established. It is easiest to consider the system in the reference frame where the nucleus is at rest. Then high energy excitations in the fast incoming proton become degenerate with the ground state. Their life-time at sufficiently high energy is significantly longer than the nuclear dimension, so  these excitations can be treated as Fock components of the proton.
In the large $N_c$ approximation one can reorganise these excitations in a series of color neutral Fock states consisting of quark-antiquark pairs \cite{Mueller:1994jq}. We consider only those Fock components of the proton, which are brought to mass shell by interactions, otherwise they remain being a virtual fluctuation of the proton.
 The $n$-th Fock state is actualized by $n$ collisions with target nucleons. Data can be explained when in $n$ collisions only $(n+1)/2$ as many particles are formed as in a single pp collision \cite{Elias:1979cp,Back:2004je,Adam:2014qja}. How can one understand the phenomenon that the fragmentation products do not multiply $n$ times, given the two facts that there are $n$ collisions and that the fragmenting two strings formed in each collision overlap strongly in rapidity space?

In this paper we present a model which can explain this phenomenon. We assume that in proton-proton collisions and in $q\bar q$-proton collisions the  strings hadronize in a similar fashion, producing a symmetric maximum entropy distribution \cite{Pirner:2011ab,Pirner:2012yy} in its respective rest frame. With the simplification that pions are produced in the same amount as gluons, the maximum entropy method yields a Bose-type distribution depending on rapidity $y$ and transverse momentum $p_{\bot}$. The maximum entropy distribution has three parameters, namely an effective transverse ``temperature'' $\lambda(s)$ and a longitudinal softness $w(s)$, both depending on the center-of-mass (cm) energy. It is symmetrical around the cm-rapidity $y_\mathrm{cm}$. Using the rapidity variable we can describe both hemispheres of the distribution:
\begin{equation}
n(y,\vec p_{\bot})=
\frac{1}{e^{\sqrt{p_{\bot}^2+m_{\pi}^2}(\frac{1}{\lambda(s)}+\frac{w(s)}{\sqrt{s}}\exp|y-y_{cm}|)} -1}.
\end{equation}

For $pp$-collisions the transverse phase space is homogeneously distributed over the area $L_{\bot}^2$. Empirical values for this area give sizes $L_{\bot} \approx 1.3$\,fm for pion distributions \cite{Pirner:2011ab,Pirner:2012yy}. Invoking parton-hadron duality the  multiplicity of produced particles is obtained by integrating the light cone distribution over the respective phase space. Note that the relativistic measure $dy= dx/x$ arises from the large spatial extension in longitudinal direction of the small $x$ partons. With a gluon degeneracy factor $g=2 (N_c^2-1)$ \cite{Pirner:2011ab} the total multiplicity becomes:
\begin{equation}
N=g \frac{N_\mathrm{part}}{2} L_{\bot}^2\int \frac{d^2 p_{\bot}}{(2 \pi)^2} \int dy \; n(y,\vec p_{\bot}).
\label{eq:multa}
\end{equation}
We emphasize that a statistical understanding of the final state in heavy-ion collisions necessitates both a correct description of the momentum and configuration space distribution. The maximum entropy model was originally conceived for symmetric pp or AA collisions. It can be adapted to the pA configuration by keeping the symmetry of the  fragmentation products with respect to the cm-rapidity which depends on the momentum of the $q \bar q$ substate in the fast proton colliding with a target nucleon.

\section{The Fock State Decomposition and multiplicity in proton-nucleus collisions}

For proton-nucleus collisions we add the individual contributions of the different projectile Fock states, which lose coherence, interacting with the nuclear participants. This means that a 2-particle Fock state is actualized (brought to mass shell) when the incoming proton interacts with two target nucleons, or a 3-body component interacts with 3 target nucleons and so on. The elastic proton-nucleus cross section amplitude, can be calculated within the Glauber theory, as well as the quasielastic and reaction cross sections (e.g. see in \cite{kps-gap}). Thus one can calculate the reaction cross section as 
\beq
\sigma^{pA}_{reac} =  \sigma^{pA}_{tot}-\sigma^{pA}_{el}-
\sigma^{pA}_{qel}=
\int d^2b \, \left[1-e^{-T_A(b) \sigma^{pN}_{in}}\right],
\label{reac}
\eeq
where the profile function of the nucleus $T_A(b)=\int_{-\infty}^{\infty} dz\,\rho_A(b,z)$ is obtained from the nuclear density integrating over $z$ direction. 
Here we employ the optical approximation and neglect the interaction range in comparison with the nuclear radius for the sake of simplicity,
more accurate, but lengthy expressions are well known \cite{kps-gap}.

Apparently the exponentials in Eq.~(\ref{reac}) can be expanded as 
$\sigma^{pA}_{reac}=\sum_n \sigma(n)$, where
\begin{equation}
\sigma(n)= \frac{1}{n!}\int d^2b \, \left[T_A(b) \sigma^{pN}_{in}\right]^n 
e^{-T_A(b) \sigma^{pN}_{in}}.
\label{sig-n}
\end{equation}

This expression is frequently misinterpreted as an $n$-fold collision of the projectile 
proton with $n$ bound nucleons. However, the proton can interact inelastically only once, the further collisions of the proton debris may occur with cross sections, quite different from $\sigma^{pN}_{in}$. Multiple inelastic interactions, involved in (\ref{sig-n}) should be interpreted as independent collisions of different constituents of the projectile Fock component inside the nucleus. However, the cross sections of those collisions are unknown, and the Glauber model alone is unable to predict them and the multiplicity distribution.

Inelastic processes are related through the unitarity relation to the forward elastic amplitude. The latter is given by the Glauber model as a sum of multiple $NN$ elastic amplitudes. However, the one-to-one correspondence 
between inelastic processes and different terms in  
the elastic amplitude is not known within the Glauber model. This problem was solved by Abramovsky, Gribov and Kancheli (AGK) \cite{Abramovsky:1973fm}, who formulated the unitarity cutting rules for the elastic amplitude. The magnitudes and signs of different multiple interaction terms in the expanded Glauber amplitude were found to be related to a sum of unitarity cuts of different number of elastic $NN$ amplitudes (Pomerons)\footnote{The $NN$ elastic amplitude, which can be called effective Pomeron, already includes multi-Pomeron terms and is subject to the AGK cutting rules. Therefore the hadron multiplicity also depends on the number of cut Pomerons, related to the projectile Fock components, interacting with the same bound nucleon. The hadron multiplicity distribution in pA collisions will still be given by the convolution of the multiplicity distribution in each $NN$ collision with the
distribution of the number of collisions. Both are controlled by the AGK rules, but with different weight factors.}
The key point of the  AGK cutting rules is the  independence of the proton multi-Pomeron coupling on the number of cut Pomerons. This important result was proven in \cite{Abramovsky:1973fm} within the old-fashioned parton model with short rapidity-range correlations.

The AGK cutting rules provide the weight factors (\ref{sig-n}) for the  inelastic $pA$ collision corresponding to $n$ cut Pomerons. Notice that the particles produced on mass shell from $n$ cut Pomerons have to share the total energy (see below), however, the proton multi-Pomeron coupling Eq.~(\ref{sig-n}) is the same as in the elastic cross section (no cut Pomerons), therefore the cross sections $\sigma^{pN}_{in}$ should be taken at the full collision energy.
At 5 TeV we use $\sigma_{in}=70$\,mb as pp-inelastic cross-section \cite {Adam:2014qja}. The sum over all $n$, i.e. over all possible collisions gives the total inelastic cross section. 

To understand the multiplicity distribution we construct a model for the $n$-th Fock state in such a way that it reproduces the general features of the phenomenology in pA collisions. The most important empirical result is the increase of the multiplicity ratio $r$ between pA and pp collisions with the mean number $\langle n \rangle$ of collisions:
\begin{equation}
r = \frac{\langle n \rangle +1 }{2}.
\label{ratio}
\end{equation}
In order to specify the details of the model, however, more and different reaction channels would have to be analyzed. We assume that the $n$-th Fock state $|n \rangle$ is a coherent superposition of $n$-particle states with different numbers of ``valence-like'' and ``sea-like'' constituents:
\begin{equation}
|v_{1},v_{2},...v_{m},s_{1},s_{2},...s_{n-m} \rangle.
\end{equation}
The valence-like constituents $v_{i}$ carry sizable momentum fractions of the proton. The sea-like constituents $s_{i}$ are very slow and do not contribute to the hadronization significantly. The first ``valence''-like constituent in the proton has the structure $|qqq \rangle$ and all others constituents are $|\bar q q \rangle$ color neutral dipoles. In line with the maximum entropy assumption we assume that in the quantum state with $m$ valence particles each particle carries the same fraction $x_m=1/m$ of the proton momentum. Because of lack of further information we assume that the individual quantum states have equal amplitudes in the superposition, in reality the individual amplitudes may depend on the reaction dynamics.  The $n$-th Fock state in the model then has the form
\begin{eqnarray}
|n \rangle=\frac{1}{\sqrt{n}}(|1,s,s,..\rangle + |1/2,1/2,s,..\rangle + |1/3,1/3,1/3,s,..\rangle + ... +|1/n,1/n,..,1/n\rangle).
\label{eq:multb}
\end{eqnarray}

We have labeled the valence-like constituents by their respective momentum fractions $x_m=1/m$. Because of the negligible role of sea substates the $m$-th component gives $m$ times the multiplicity of a proton-proton collision, modified, of course, according to the momentum fractions  of the color-neutral states  in the Fock state. To obtain the total multiplicity $N(n)$ generated by the $n$-th Fock state we average the sum of the multiplicities of the $n$ components:
\begin{equation}
N(n)=\frac{1}{n} \frac{n (n+1)}{2}
\label{eq:mult2}
\end{equation}
Combining this multiplicity with the Glauber weights gives the empirical multiplicity ratio between pA- and pp-collisions, Eq.~(\ref{ratio}).
This simple additivity is broken by energy conservation and rapidity shifts arising from the lower momenta in the substates. There have been more sophisticated proposals \cite{Capella:1992yb} about the distribution function of the different Fock states and how they interact with the target nucleons, but the necessary multi-parton distributions are really unknown and can only be guessed.

The eikonal formula still holds because Gribov inelastic shadowing corrections to the total inelastic cross section from excitations of the fast proton are known to be small \cite{Kopeliovich:2016jjx}. It also has been shown in string motivated models \cite{Shoshi:2002fq} that the inelastic pp cross section is mainly given by the distance $(R)$ of q and diquark in the proton, whereas the string thickness ($a$) gives only a small contribution. Consequently also the transverse string excitations only give a minor modification of the cross section since the transverse size a is much smaller than the extension of the string $a \ll R$.

\section{Hadron rapidity distributions in pA-collisions}
Because of  momentum sharing the $m$-th component of the Fock state creates a multiplicity distribution with a reduced  cm-energy and a cm-rapidity shifted from the pp-cm momentum $y=0$ towards positive rapidities, i.e. towards the target rapidity in our convention. Note different experiments have different conventions for the rapidity of the nucleus and the proton, we use the one in the ALICE publication \cite{ALICE:2012xs}. Since the lead beam has an energy of $1.58$ TeV/per nucleon and the opposing proton beam an energy of $4$ TeV there is an additional displacement $\Delta$ of the cm-rapidity in proton direction:
\begin{eqnarray}
s(m)&=& x_m s\\
y_{cm}(m)&=& - \log(x_m)/2 - \Delta\\
\Delta&=&0.465.
\end{eqnarray}

Each collision between a proton substate with a target proton leads to a fragment distribution like in a pp collision. The existing parameterizations of the pp data with the maximum entropy distribution allow us to interpolate the energy dependences of the two parameters of the light cone plasma distributions. The fitted values of the parameters for cm-energies $0.2$ TeV$<\sqrt{s}<7.0$ TeV serve as input for the energy dependence of the effective
transverse temperature $\lambda(s)$ and the longitudinal softness $w(s)$:
\begin{eqnarray}
\lambda(s)&=& (0.023 + 0.03 \log(\sqrt{s/s_0})) \, \mathrm{GeV}\\
w(s) &=&-3.66+1.33 \log(\sqrt{s/s_0})\\
s_0 &=& 1 \, \mathrm{GeV^2}.
\end{eqnarray}

Additional constants in the maximum entropy distribution are $K=0.3$ and the pion mass $m_{\pi}=0.138$ GeV. The starting values $w=7.67$ and $\lambda=0.274$ GeV correspond to the cm energy  $\sqrt{s}$=5 TeV, but in the higher Fock states this energy is reduced. We calculate the relevant rapidity distribution for a collision of a substate with $x_m=1/m$ with a target proton using the correct energy $s(m)$ and rapidity $y_{cm}(m)$ to get the pseudorapidity distribution $\frac{dN(m)}{d^2p_{\bot}d\eta}$ which is the basic building block of our model:
\begin{eqnarray}
\frac{dN(m)}{d^2p_{\bot}d\eta} &=& \frac{16 L_{\bot}^2}{(2\pi)^2}\sqrt{1-\frac{m_{\pi}^2}{(m_{\pi}^2+p_{\bot}^2) \cosh^2y(\eta,p_\perp)}} n_m(y,p_{\bot})\\
 n_m(y,p_{\bot})&=& \left(e^{\sqrt{p_{\bot}^2+m_{\pi}^2}\langle{\frac{1}{\lambda(s(m))}+
\frac{w[s(m)]}{K\sqrt{s(m)}}\exp|y(\eta,p_\perp)-y_{cm}(m)|}\rangle} -1 \right)^{-1}\\
y(\eta,p_\perp)&=&\log\left[\frac{\sqrt{m_{\pi}^2+p_\perp^2 \cosh^2\eta}+p_\perp \sinh\eta}{\sqrt{m_{\pi}^2+p_\perp^2}}\right].
\end{eqnarray}

\begin{figure}
\centering
\includegraphics[width=0.7\linewidth]{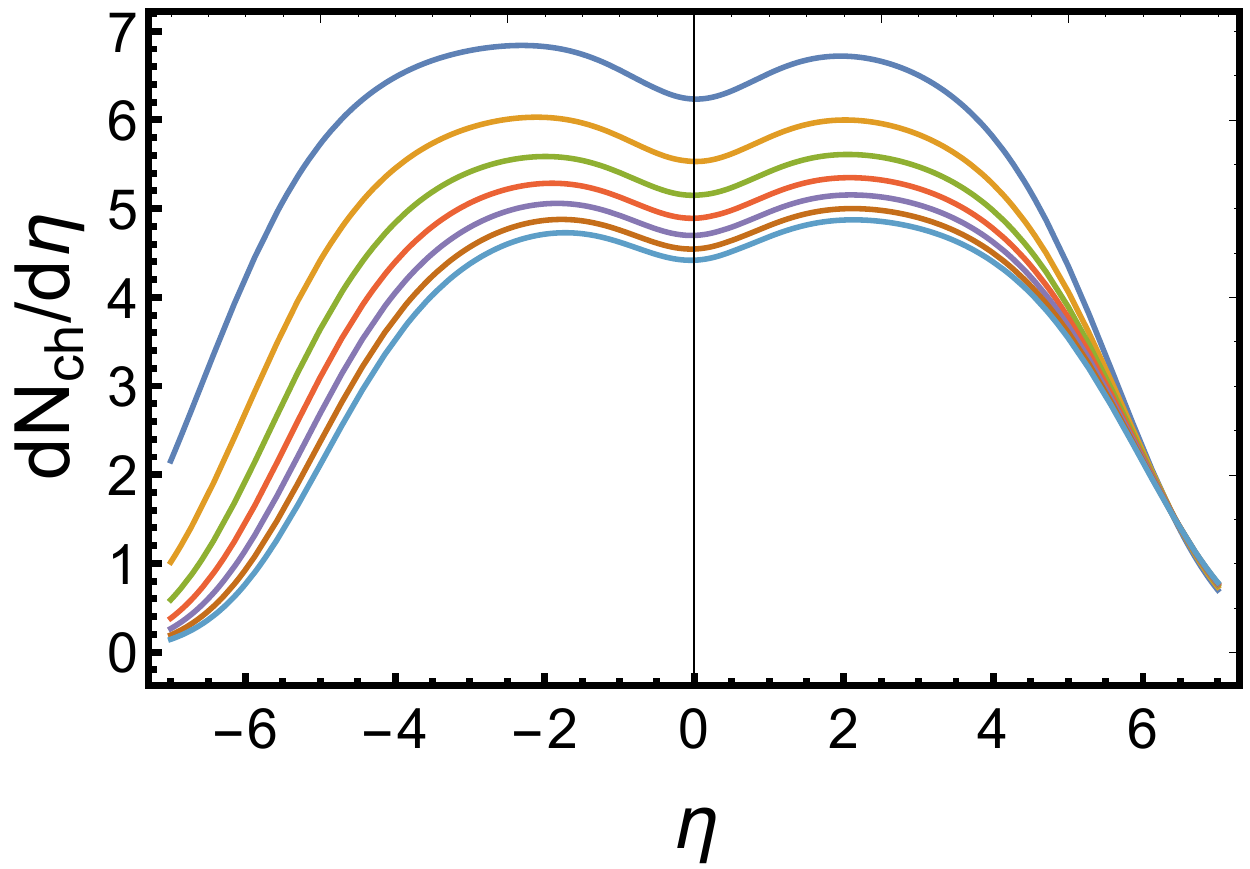}
\caption{Individual contributions $\frac{dN_{ch}(m)}{d\eta}$ with $m=1$ to $m=7$ from top to bottom are shown for $\sqrt{s}=5$\,TeV. The cm-momenta are shifted towards the target rapidity and cm-energies are reduced due to momentum sharing.}
\label{fig:individual}
\end{figure}

The $n$-th Fock state $|n\rangle$ interacting with $n$ target nucleon can produce up to $n$ times the proton-proton multiplicity with the probability 
\beq
P(n) =\frac{\sigma(n)}{\sigma_{reac}}.
\label{probab}
\eeq
The first component $|1,s,s,s,...\rangle$ gives a rapidity distribution $\frac{dN(1)}{d^2p_{\bot}d\eta}$. The second component $|1/2,1/2,s,s,... \rangle$ gives twice the rapidity distribution $\frac{dN(2)}{d^2p_{\bot}d\eta}$ and similarly for the higher components. The total sum of produced fragmentation products comes from the arithmetic sum of $1+2+3+...+n$ pp distributions which must be averaged, because each component has equal weight.  Averaging gives the correct  multiplicity ratio between pp and pA collisions observed for high energy collisions as discussed before. For the charged particle multiplicity we multiply with a factor $2/3$:

\begin{equation}
\frac{dN_{ch}}{d^2p_{\bot}d\eta} = \frac{2}{3} 
\sum_{n=1}^{n_{max}} P(n)\frac{1}{n} \left( \frac{dN(1)}{d^2p_{\bot}d\eta}+2 \frac{dN(2)}{d^2p_{\bot}d\eta} + 3 \frac{dN(3)}{d^2p_{\bot}d\eta}+ ... + n \frac{dN(n)}{d^2p_{\bot}d\eta} \right)
\label{pT}
\end{equation}
The summation is extended to $n_{max}$=40, since  higher contribution are negligible. 
To compare with the data we integrate over transverse momentum:
\begin{eqnarray}
\frac{dN_{ch}}{d\eta} &=& \int d^2p_{\bot} \, \frac{dN_{ch}}{d^2p_{\bot}d\eta},\\
 &=& \sum_{n=1}^{n_{max}} P(n)\frac{1}{n} \left( \frac{dN_{ch}(1)}{d\eta}+2 \frac{dN_{ch}(2)}{d\eta} + 3 \frac{dN_{ch}(3)}{d\eta}+ ... + n \frac{dN_{ch}(n)}{d\eta} \right).
\end{eqnarray}

\begin{figure}
\centering
\includegraphics[width=0.7\linewidth]{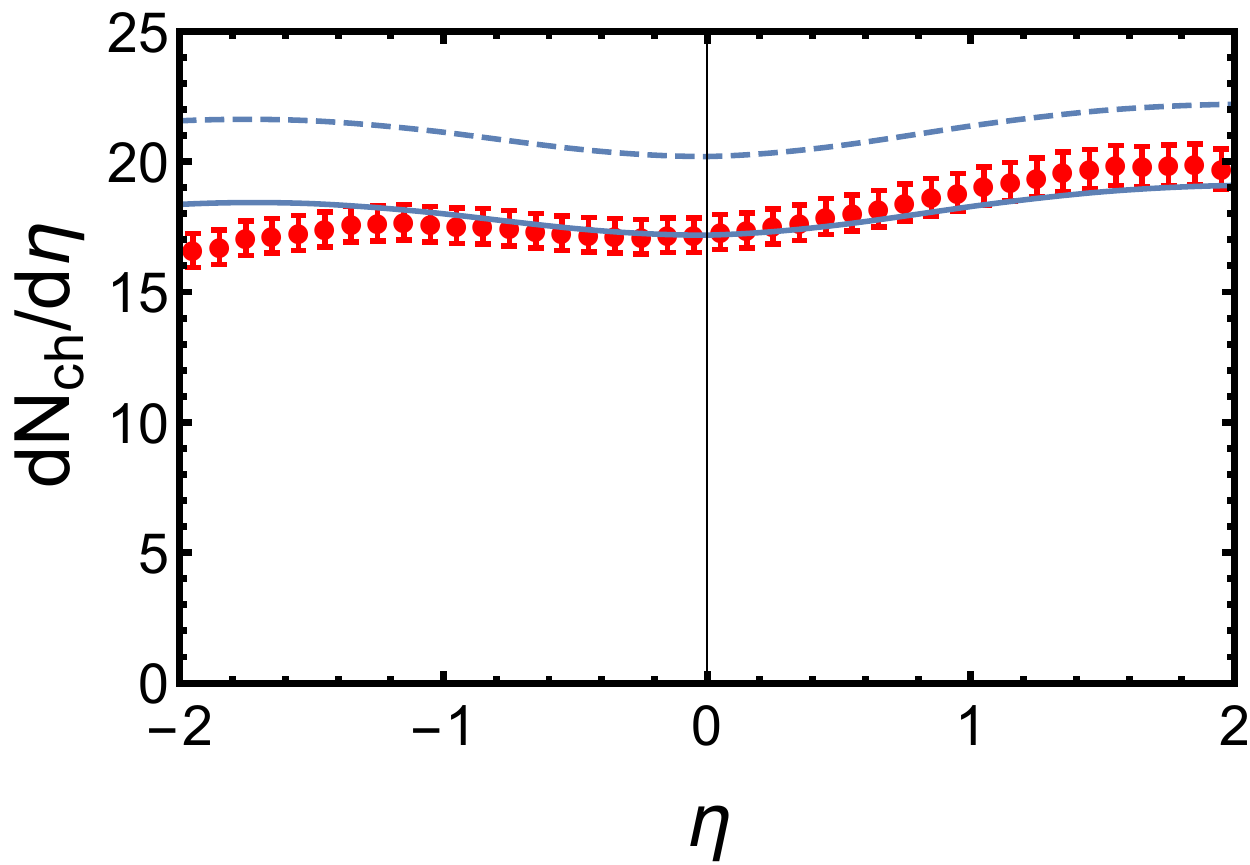}
\caption{Data points \cite{ALICE:2012xs} from the ALICE experiment show the charged-particle pseudorapidity distribution in p+Pb collisions at  $\sqrt{s} = \unit[5020]{GeV}$, the solid curve represents the result from the presented model. The dashed curve gives our prediction for the 8000 GeV data.}
\label{fig:pPb}
\end{figure}

In Fig.~\ref{fig:individual} the contributions $\frac{dN_{ch}(m)}{d\eta}$ for $m=1$ to $m=7$ are shown individually. Due to momentum sharing they decrease in magnitude and move towards positive rapidities. For p-Pb collisions on the average $\langle n \rangle = 7.25$ collisions occur. Summing over all collisions one obtains the p-Pb rapidity distribution in Fig.~\ref{fig:pPb}. The lower curve shows the 5 TeV data from the ALICE collaboration \cite{ALICE:2012xs} together with the theory. The theoretical calculation is close to the data, but slightly underestimates the asymmetry of the measured distribution in pseudo-rapidity. The neglected small sea substates in the Fock states would increase the theoretical result for positive rapidities and decrease the proton side.

A comparison of our theoretical prediction with the data of 8 TeV p-Pb collisions may present an additional test of our model. The theoretical multiplicity distribution at 8 TeV has the same shape as at 5 TeV, but one obtains a larger multiplicity due to the higher energy.

Notice that $p_{\bot}$-dependence Eq.~(\ref{pT}) does not exhibit the expected effect of broadening. This is not the problem of the Fock-state picture, but of the Glauber approximation. The unitarity cut of the Glauber elastic pA amplitude
suggested by the AGK rules, assumes independent hadronization
of the cut Pomerons. This assumption, known as Bethe-Heitler approximation, is subject to corrections for the effects of coherence, which is known as the effect of saturation. Namely, a hadron can be produced coherently by several cut Pomerons, leading to rising $\la p_{\bot}^2\ra$. Numerically this is a rather weak effect, as was evaluated in \cite{kps-broad}.
Even weaker is its influence on the mean $p_{\bot}$-integrated multiplicity,
so we neglect this correction. Smallness of this correction is confirmed by successful calculations of the multiplicity distribution in pA collisions performed with the Glauber Monte-Carlo \cite{klaus}.

\section{The Hadron Rapidity distribution in high multiplicity proton-nucleus collisions}

\begin{figure}
\centering
\includegraphics[width=0.7\linewidth]{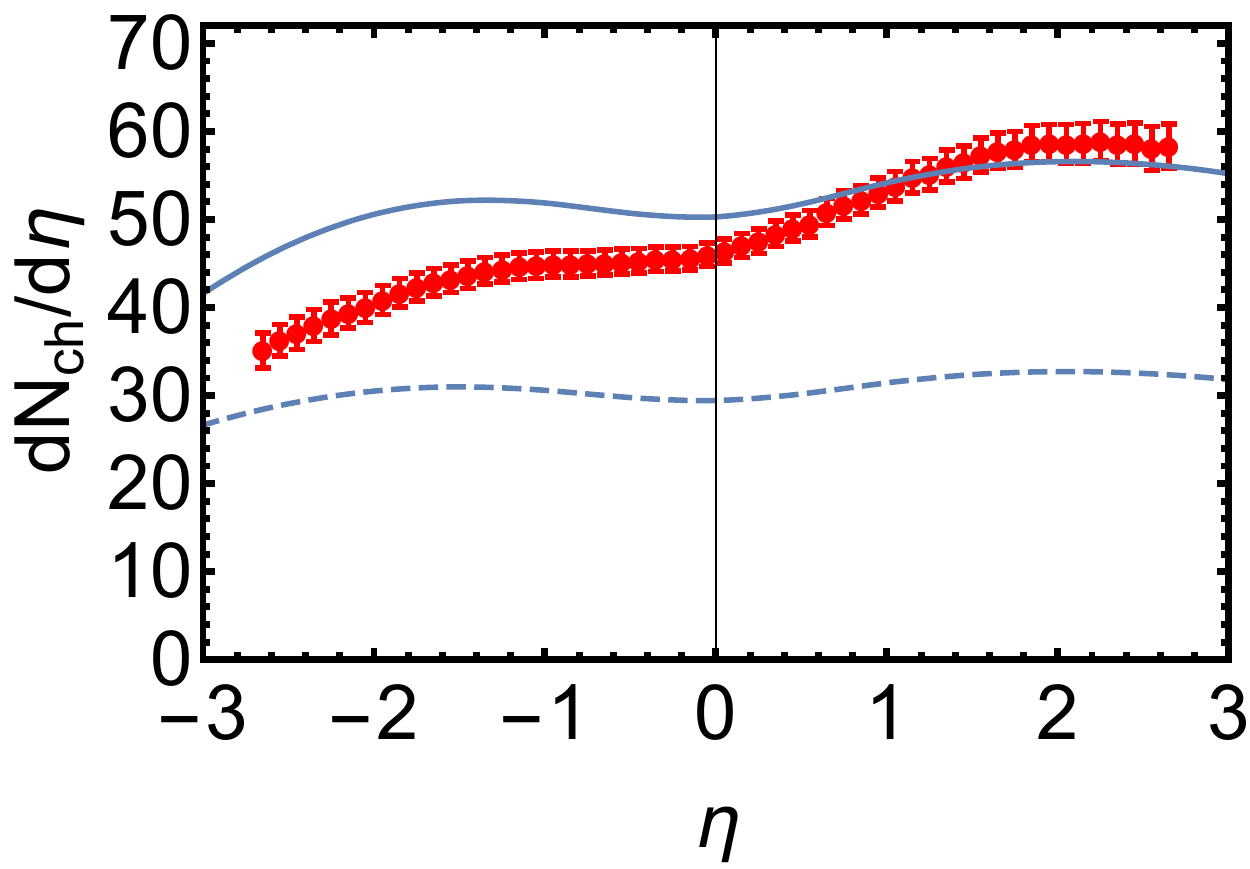}
\caption{Data points from the ATLAS experiment \cite{Aad:2015zza} show the charged-particle pseudorapidity distribution for the 5\% most central p-Pb collisions at $\sqrt{s}= \unit[5020]{GeV}$, the solid curve represents the result from the presented model selecting the highest component in each Fock component. The dashed curve gives the prediction for the averaged Fock states.}
\label{fig:pPbhigh}
\end{figure}

The highest multiplicity p-Pb data show a surprisingly large enhancement of the rapidity distribution, especially in the target hemisphere. We remind that the minimum bias ratio  of pA to pp multiplicity is given by Eq.~(\ref{ratio}).
A high multiplicity trigger does select the highest component of each Fock-state $|n,\mathrm{high} \rangle$ in our model, namely
\begin{equation}
|n,\mathrm{high} \rangle=|1/n,1/n,..,1/n\rangle
\label{eq:mult}
\end{equation}
which produces $n$ times the proton multiplicity and consequently will increase the overall multiplicity by a factor two:
\begin{equation}
r_\mathrm{high}= \langle n \rangle.
\end{equation}
In addition the selection of impact parameter will also give a higher multiplicity. In Fig.~\ref{fig:pPbhigh} we show the highest  multiplicity we can get in our model for an impact parameter cut $0 < b <1.76\,\mathrm{fm}$ which corresponds to a centrality of $0$--$5$\%.

It is impossible to get the factor three enhancement only from a cut in impact parameter space. In our  model the average Fock state (cf.\ Fig.~\ref{fig:pPbhigh}) does not suffice to describe the data taken in the ATLAS experiment \cite{Aad:2015zza}, whereas the selected Fock state gives  good agreement in the positive rapidity region. There is an even stronger asymmetry in the data than in Fig.~\ref{fig:pPb} which is missed by the simplified theory. Possible final state effects can also change our results.

In a recent reference \cite {Albacete:2017qng} predictions for the 8 TeV p+Pb run at the LHC are compiled and compared with each other. They include saturation approaches, Monte Carlo event generators and perturbative  QCD-based calculations. In contrast to these basic approaches, which aim at an ab initio description of pp- collisions, our paper uses the phenomenological rapdity distribution of pp-collisions to generate the rapidity distribution in pA-collisions. Using the maximum entropy distribution for pp-collisions, we  can with the same ansatz obtain the rapidity distribution in the most and less central regions. We develop a new view on the fluctuations in the proton, which extends previous work \cite{Dusling:2017aot} to study fluctuations in the proton wave function explaining the observed azimuthal asymmetries of produced particles. A study of color neutral components in the proton is mostly known from the pion cloud picture. One can interpret the valence $q \bar{q}$ states as the high energy manifestation of the pion cloud, its chiral partner and other mesons. In the fast proton the container and cloud of the quarks carry quanta with sizeable light cone momenta which can interact with the target nucleons. This is the main new ingredient of our presented model.

Several improvements have to be studied: (i) How does the focusing on the highest Fock state takes place as a function of the trigger in order to understand the variations of the rapidity distribution with centrality.
(ii) How do the sea states enter the Fock state composition quantitatively?
(iii) Can the maximum entropy distribution for string fragmentation fluctuate?
In this paper we only use an average value for the transverse temperature $\lambda$, but one expects that fluctuations in $\lambda$ may be important to boost the multiplicity which is proportional to $\lambda^2$. In statistical physics it is well known that fluctuations of the ``temperature'' can lead to a power behavior of the high $p_{\bot}$ part of the transverse momentum distribution which is encoded in Tsallis distributions. (iv) New experimental tests could be pp- and AA-collisions analyzed along the same lines as outlined in the paper. 
These points will be studied in further work. 

\section*{Acknowledgements}
We thank Klaus Reygers for discussions and critical reading of the manuscript. This work was supported in part by Fondecyt grants 1170319 and 1140377 (Chile), by Proyecto Basal FB 0821 (Chile), and by Conicyt grant  PIA ACT1406 (Chile).









%

\end{document}